\documentclass[pdftex]{article}
\pdfoutput=1
\usepackage{amssymb}
\usepackage{icrctc07}

\title{Galactic neutrino background from cosmic ray interaction with
  the ISM content}
\shorttitle{Galactic neutrino}

\authors{C. De Donato$^{1}$ , G. A. Medina-Tanco$^{2}$, J. C.  D'Olivo$^{2}$}

\shortauthors{De Donato C. et al.}

\afiliations{
$^1$Dipartimento di Fisica dell'Universit\`a degli Studi di Milano and INFN, Milano, Italy, \\
$^2$Dep. Altas Energias, Inst. de Ciencias Nucleares, Universidad
Nacional Autonoma de Mexico, Mexico DF, CP 04510}
\email{cinzia.dedonato@mi.infn.it}

\abstract{
We use a diffusive model for the propagation of Galactic cosmic rays to
estimate the
charged pion production in interactions with protons of the interstellar
medium.
Cosmic ray nuclei from proton to iron are considered and the corresponding
contribution to the neutrino secondary flux produced as a result of
spallation is
also estimated. 
}

\begin{document}
\maketitle

\section{Introduction}
We used the numerical Galactic Cosmic Ray  propagation code GALPROP
\begin {footnote}{$http://galprop.stanford.edu/web\_galprop/\\galprop\_home.html$}\end{footnote} \cite{Strong1998} to reproduce the
diffusive galactic spectrum from SuperNova Remnants (SNRs) and estimate  the
charged pion production in interactions with protons of the interstellar medium.
From this estimation we calculate the flux of neutrinos coming from the decay of charged pions.\\
\section{Diffusion Galactic model}
\label{DiffusionGalacticModel}
The diffusive model assumes cylindrical symmetry in the Galaxy, with coordinates
R and z equal to the Galactocentric radius and the distance from the Galactic plane.
The propagation region is bounded  by $R=R_h=30~kpc$ and $z=z_h=4~kpc$, beyond which free escape is assumed.
The diffusion coefficient is taken as $\beta D_0(\rho / \rho_D)^{\delta}$ , where $\rho$ is the particle rigidity, 
$D_0$ is the diffusion coefficient at a reference rigidity  $\rho_D$ and $\delta=0.6$.\\
The distribution of cosmic rays sources used reproduces the cosmic-ray distribution determined by the analysis
of EGRET gamma-ray data \cite{Strong1998}.

Nuclei with $Z<26$ are injected with a rigidity power law spectrum of index $\alpha=-2.05$,
 independently of energy, with isotopic abundances given by space measurements of
the cosmic ray abundances \cite{Strong2001}.
The neutrino flux is calculated as the product of the decay of charged pions  
which, in turn, are created in collisions
of cosmic-ray particles with interstellar gas.\\
The used interstellar  molecular, atomic and ionized, (H2 , HI, HII) hydrogen distribution are derived from radio
 HI and CO surveys in 9 Galactocentric rings and from  information on the ionized component.
The distribution of molecular hydrogen is derived indirectly from CO radio-emission and the assumption that the conversion
factor H2/CO is the same for the whole Galaxy \cite{StrongMattox1996}.
 The atomic hydrogen (HI) distribution is taken from \cite{GordonBurton1976a}, with a z-dependence
calculated using  two approximation at different galactocentric distances R \cite{DickeyLockman1990,Cox1986}
 The ionized component HII is calculated using a cylindrically symmetric model \cite{Cordes1991}.

\section{Pion production}
Pion production  in $pp$-collisions is calculated following a method developed by Dermer, which combines isobaric
and scaling models of the reaction \cite{MoskalenkoStrong1998}.
The two models work well at low and high energy
respectively. In the energy range $3~GeV-7~GeV$ an interpolation of the two models is used.\\
In the isobaric model the distribution of pions is calculated by the integration over the isobar mass spectrum ($M_1=m_p+ m_\pi$, $M_2=\sqrt{s}-m_p$):
{\setlength\arraycolsep{0pt}
\begin{eqnarray}
&&F_\pi(E_\pi, E_p) = \int_{M_1}^{M_2} dm_\Delta \frac{f_{\pi}(E_\pi, E_p;m_\Delta)}{(m_\Delta - m_\Delta^0)^2+ \Gamma^2}\nonumber\\
&&\times\frac{\Gamma }{tan^{-1}\left( \frac{M_2-m_\Delta^0}{\Gamma}\right)-
tan^{-1}\left( \frac{M_1 -m_\Delta^0}{\Gamma}\right)  }
\end{eqnarray}
}\noindent
where $E_\pi$ and $E_p$ are the pion and proton energy in the laboratory system (LS), $m_\Delta^0$ is the average mass of the
 $\Delta$-isobar, $\Gamma$ is the width of the Breight-Wigner distribution and $\sqrt{s}$ is the CMS energy.
In the model it is assumed that the produced $\Delta$-isobar of mass $m_\Delta$ has either the same direction (+)
or the opposite direction (-) of the colliding proton in the CMS.
The produced isobar decays isotropically
producing a pion with the distribution:
{\setlength\arraycolsep{0pt}
\begin{eqnarray}
&&f_{\pi}(E_\pi, E_p;m_\Delta)= \frac{1}{4m_\pi \gamma_\pi ' \beta_\pi'} \times\\
&&\bigg\lbrace
\frac{1}{\gamma_\Delta^+\beta_\Delta^+}H[\gamma_\pi; a^+,b^+]+ \frac{1}{\gamma_\Delta^-\beta_\Delta^-}H[\gamma_\pi;a^-,b^-]
\bigg\rbrace \nonumber
\end{eqnarray}
}\noindent
where  $H[x;a,b]=1$ if $a\leq x \leq b$ and $H[x;a,b]=0$ otherwise,
with $a^\pm = $$\gamma_\Delta^\pm \gamma_\pi'(1-\beta_\Delta^\pm\beta_\pi')$ and
$b^\pm= \gamma_\Delta^\pm\gamma_\pi'(1+\beta_\Delta^\pm\beta_\pi')$
The Lorentz factors of the forward (+) and backward (-) moving isobars are
$\gamma_\Delta^{\pm} = \gamma_c\gamma_\Delta^* (1\pm \beta_c\beta_\Delta^*)$
 where $\gamma_c= \sqrt{s}/2m_p$ is the Lorentz factor of the CMS in the LS and
$\gamma_\Delta^*=(s+m_\Delta^2-m_p)/2\sqrt{s}m_\Delta$  is the Lorentz factor of the isobar in the CMS.
The pion Lorentz factor in the rest frame of the $\Delta$-isobar is $\gamma_\pi'= (m_\Delta^2+m_\pi^2-m_p^2)/2m_\Delta m_\pi$.\\

The scaling model gives the lorentz invariant cross section for pion production as:
\begin{eqnarray}
E_\pi \frac{d^3\sigma}{d^3p_\pi}&=& AG_\pi(E_p)(1-\tilde{x}_\pi)^Q\\
&\times& exp\left[-\frac{Bp_\bot}{1+4m_p^2/s}\right],\nonumber
\end{eqnarray}\noindent
where
{\setlength\arraycolsep{0pt}
\begin{eqnarray}
&&G_{\pi^\pm}(E_p)= (1+4m_p^2/s)^{-R},\\
&&G_{\pi^0}(E_p)=(1+23 E_p^{-2.6})(1-4m_p^2/s)^R,\\
&&Q=(C_1-C_2p_\bot+C_3p_\bot^2)/\sqrt{1+4m_p^2/s},\\
&&\tilde{x}_\pi= \sqrt{x^*_\parallel + (4/s)(p_\bot^2+m_\pi}),\\
&&x^*_\parallel= \frac{2m_\pi\sqrt{s}\gamma_c\gamma_\pi(\beta_\pi cos\theta-\beta_c)}{[(s-m_\pi^2-m_X^2)^2-4m_\pi^2m_X^2]^{1/2}},
\end{eqnarray}
}\noindent
$\theta$ is the pion polar angle in LS, $A,~B,~C_{1,2,3},~R$ are positive constants and
$m_X$ is the X channel of the reaction ($pp\rightarrow \pi^\pm + X$).\\
The energy distribution of pions can be obtained integrating over the polar angle $\theta$
\begin{eqnarray}
F_\pi(E_\pi,E_p)&=& \frac{2\pi p_\pi}{\langle\eta\sigma(E_p)\rangle_{sm}}\\
&\times&\int_{cos\theta_{min}}^{1}dcos\theta\left(E_\pi \frac{d^3\sigma}{d^3p_\pi}\right),\nonumber
\end{eqnarray}\noindent
where
\begin{eqnarray}
cos\theta_{min}= \frac{\left( \gamma_cE_\pi- \frac{s-m_X^2+m_\pi^2}{2\sqrt{s}}\right)}{\beta_c\gamma_cp_\pi}.
\end{eqnarray}

The distribution of muon neutrinos produced directly by the  decay of pions (or kaons) produced in a $pp$-collision
is given by 
{\setlength\arraycolsep{0pt}
\begin{eqnarray}
&&F(E_\nu, E_p)=\int_{E_{\pi}^{min}}^{E_{\pi}^{max}}\!\!\!\!\!\!\!\! dE_{\pi}  F_{\pi}(E_\pi,E_p) \frac{dn}{dE_\nu}=\\
&&=\int_{E_{\pi}^{min}}^{E_{\pi}^{max}} dE_{\pi} F_{\pi}(E_\pi,E_p) \frac{BR}{(1-m_\mu^2/M_\pi^2)P_\pi},\nonumber
\end{eqnarray}
}\noindent
with $E_{\pi}^{min}= E_\nu/(1-r) + M_\pi^2(1-r)/(4 E_\nu)$,  $E_{\pi}^{max}=(s-M_X^2+M_\pi^2)/(2 \sqrt{s})$, 
$BR$ is the branching ratio for meson decay in muons and $r=m_\mu^2/M_\pi^2$.

\section{Muon decay}
Since the muons originated from pions are produced fully polarized,
the energy distribution of the neutrinos/antineutrinos in the muon rest frame is given by
\begin{equation}
\frac{dn}{dxd\Omega}= \frac{1}{4\pi}[f_0(x)\mp f_1(x)cos\theta]
\end{equation}\noindent
where $x=E_\nu'/m_\mu$  with $E_\nu'$ the neutrino energy in the muon rest frame and $\theta$
the polar angle between the neutrino and the muon spin.
The functions $f_0(x)$ and $f_1(x)$  are given in table \ref{F01}.
\begin{table}[t]
\begin{center}
\begin{tabular}{|l|c|c|}
\hline
  & $f_0(x)$ & $f_1(x)$  \\
\hline
$\nu_\mu$  & $2x^2(3-2x)$ & $2x^2(1-2x)$  \\
$\nu_e$   & $12x^2(1-x)$& $12x^2(1-x)$  \\
\hline
\end{tabular}
 \caption{Functions for neutrinos from muon decay}
\label{F01}
\end{center}
\end{table}

Integrating over the polar angle and trasforming the distribution to the LS,
the energy distribution of a neutrino in the LS becomes:
\begin{equation}
\frac{dn}{dy}= \frac{1}{\beta_\mu}[g_0(y,\beta_\mu)- Pol_\mu g_1(y,\beta_\mu)]
\end{equation}\noindent
where $y=E_\nu/E_\mu$, $E_\mu$ is  the muon energy and momentu in the LS and $Pol_\mu$ is the muon polarization. 
The functions $g_0(y,\beta_\mu),~g_1(y,\beta_\mu)$  are given by
{\setlength\arraycolsep{1pt}
\begin{eqnarray}
g_0(y, \beta_\mu)&=&\int_{x_{min}}^{x_{max}}f_0(x) \frac{dx}{x},\\
g_1(y, \beta_\mu)&=&\int_{x_{min}}^{x_{max}}f_1(x)\frac{2y/x-1}{\beta_\mu} \frac{dx}{x}
\end{eqnarray}
}\noindent
with $x_{min}=2y/(1+\beta_\mu)$ and  $x_{max}=min[1,2y/(1-\beta_\mu)]$.\\

\section{Neutrino production}

The distributions of muons from the decay of mesons (pions or kaons) is given by
\begin{equation}
\frac{dn}{dE_\nu}= \frac{dn}{dE_\mu}= \frac{BR}{(1-m_\mu^2/M^2)P_M}
\end{equation}\noindent
where $M$ is the pion/kaon mass, $P_M$ is the pion/kaon momentum in the LS and $BR$ is the branching ratio of the decay.\\
The energy distributions of neutrinos  from muons through the pion (or kaon) decay in the LS is given by:
{\setlength\arraycolsep{1pt}
\begin{eqnarray}
F(E_\nu, E_\pi)&=& \int_{E_{\mu}^{min}}^{E_{\mu}^{max}}dE_\mu \frac{dn}{dE_\mu} \frac{dn}{dE_\nu}=\nonumber\\
&=& \int_{y_{min}}^{y_{max}} dy \frac{dn}{dE_\mu}\frac{1}{y} \frac{dn}{dy},
\end{eqnarray}
}\noindent
where $y=E_\nu/E_{\mu}$, $y_{min}=E_\nu/E_{\mu}^{max}$, $y_{max}=E_\nu/E_{\mu}^{min}$ and 
{\setlength\arraycolsep{1pt}
\begin{eqnarray}
 E_{\mu}^{min}&=& max[m_\mu, \gamma_\pi(E_\mu^*-\beta_\pi P_\mu^*) ],\\
 E_{\mu}^{max}&=& max[m_\mu, \gamma_\pi(E_\mu^*+\beta_\pi P_\mu^*) ],
\end{eqnarray}
}\noindent
with
$\gamma_\pi$ the pion Lorentz factor in the LS and $E_\mu^*,~p_\mu^*$ the muon energy and momentum in the pion rest frame:
\begin{eqnarray}
E_\mu^*&=&\frac{M_\pi^2+m_\mu^2}{2M\pi},\\
P_\mu^*&=&\frac{M_\pi^2-m_\mu^2}{2M\pi}.
\end{eqnarray}

Taking into account pion production through $pp$-collisions we have:
{\setlength\arraycolsep{1pt}
\begin{eqnarray}
F(E_\nu, E_p)&=&\int_{E_{\pi}^{min}}^{E_{\pi}^{max}} dE_\pi\\
& &\int_{y_{min}}^{y_{max}} dy F_\pi(E_\pi,E_p)\frac{dn}{dE_\mu}\frac{1}{y} \frac{dn}{dy},\nonumber
\end{eqnarray}
}\noindent
with 
\begin{eqnarray}
E_{\pi}^{min}&=&M_\pi \times\\
&& max\left[1, \frac{E_\nu}{E_\mu^*+P_\mu^*} + \frac{E_\mu^*+P_\mu^*}{4E_\nu} \right]\nonumber\\
E_{\pi}^{max}&=&\frac{s-M_X^2+M_\pi^2}{2 \sqrt{s}}. 
\end{eqnarray}

\begin{figure}[!ht]
\begin{center}
\includegraphics [height= 9.0cm, width=0.48\textwidth]{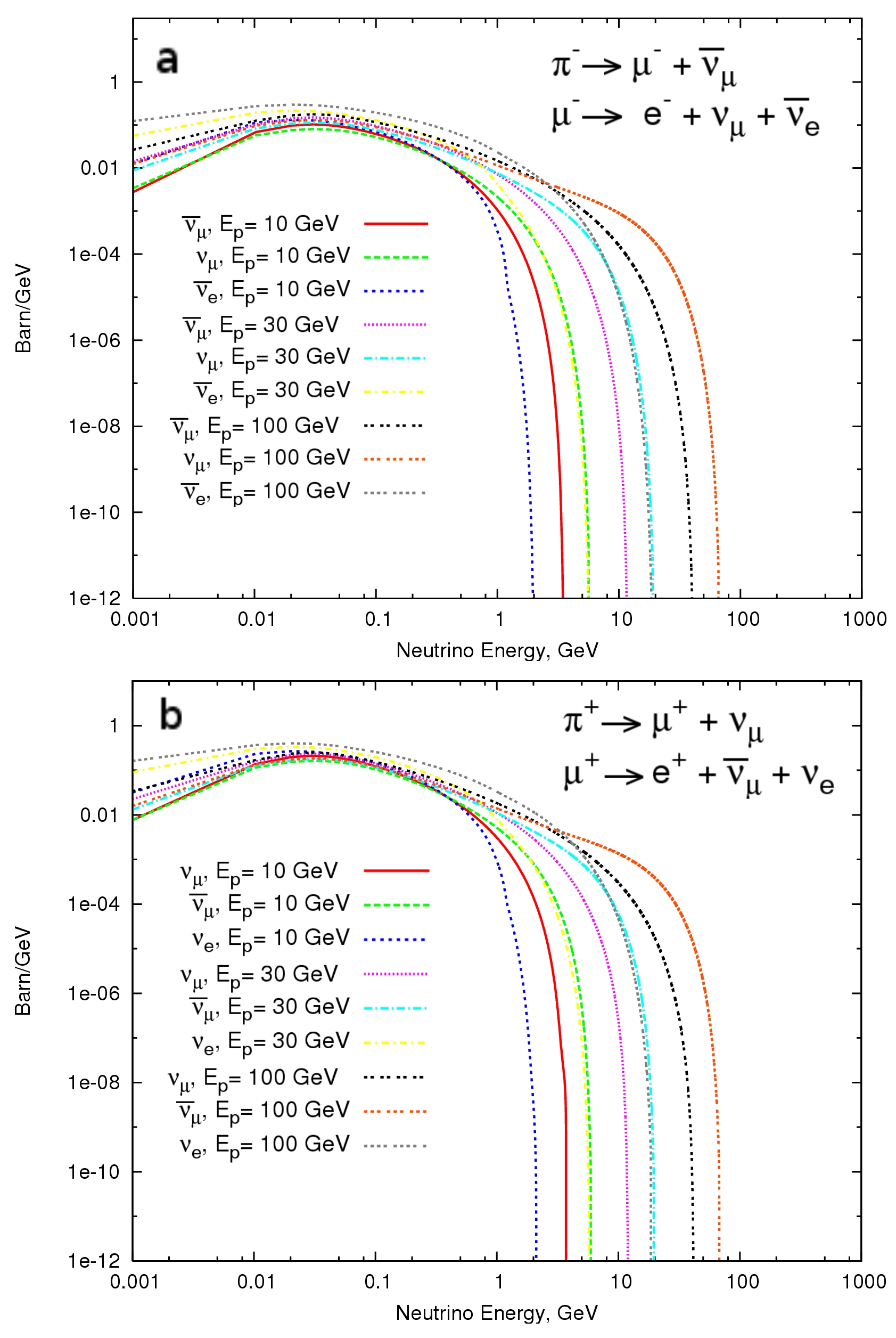}
\end{center}
\caption{Calculated cross section for neutrinos  from negative (a) and positive (b) pions decay  produced in $pp$-collisions.
The cross section has been calculated for fixed value of the cosmic ray proton energy.}\label{fig01}
\end{figure}

Fig.\ref{fig01} shows the neutrino production cross section from the decay of negative and positive pions.  
For a given proton energy, we calculated the cross section for the muon neutrinos produced directly from the pion decay and 
for the electron and muon neutrinos produced from the muon decay. 

\section{Neutrino flux}

\begin{figure}[!ht]
\begin{center}
\includegraphics [height= 5.7cm, width=0.48\textwidth]{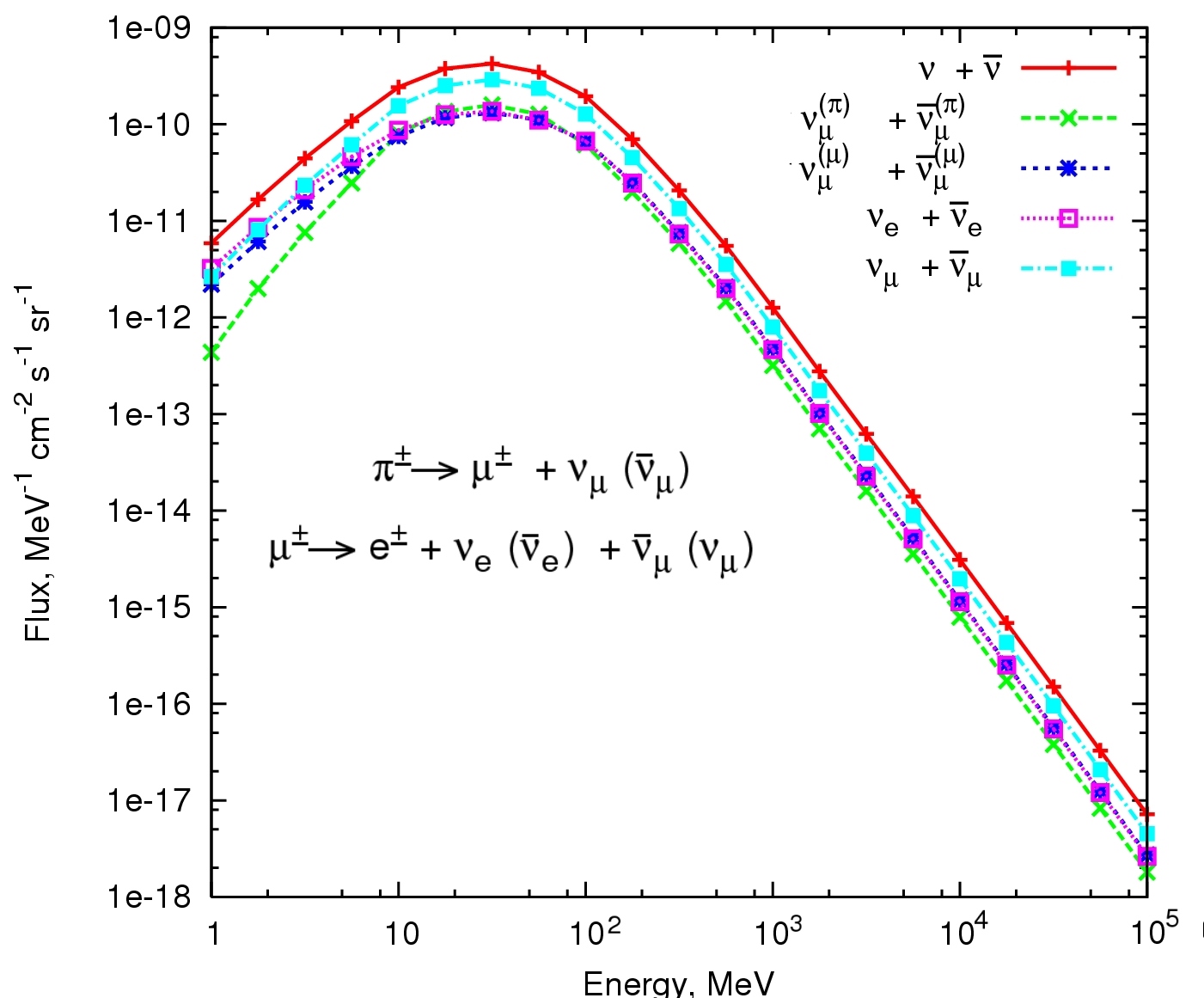}
\end{center}
\caption{Flux of neutrinos coming from the Galactic bulge produced by pion decay through the interaction of cosmic rays with the ISM gas. 
The curves indicates the total neutrino/antineutrino flux ($\nu + \overline{\nu}$), the electron ($\nu_e + \overline{\nu}_e$) 
and muon ($\nu_\mu + \overline{\nu}_\mu$) contributions and the different muon neutrino components, 
neutrinos coming directly from the pion decay ($\nu_\mu^{(\pi)} + \overline{\nu}_\mu^{(\pi)}$) 
and neutrinos coming from the muon decay ($\nu_\mu^{(\mu)} + \overline{\nu}_\mu^{(\mu)}$).}
\label{fig02}
\end{figure}

\begin{figure}[!ht]
\begin{center}
\includegraphics [width=0.48\textwidth]{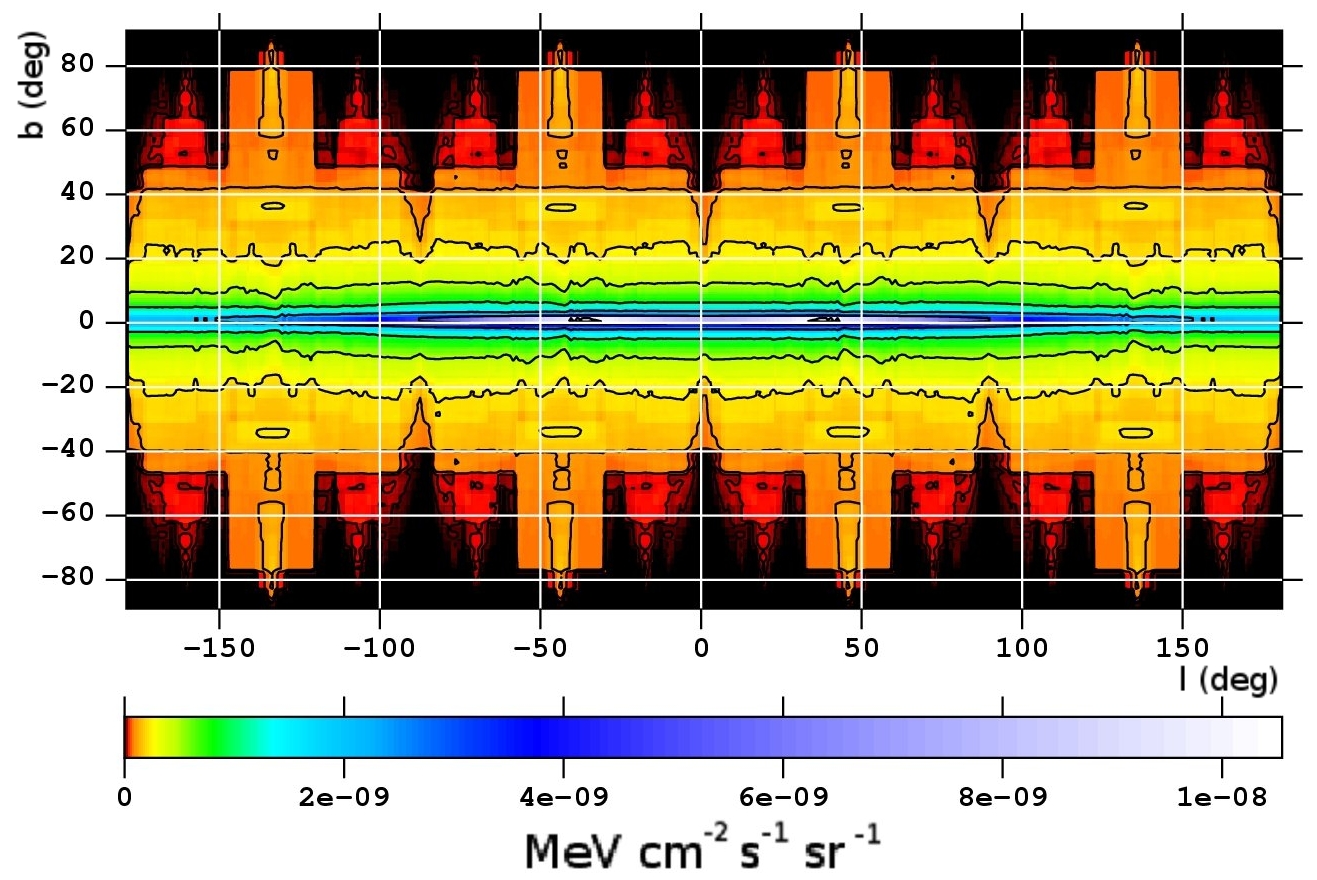}
\end{center}
\caption{Sky map of the total neutrino energy flux produced by pion decay through the interaction of galactic cosmic rays with the ISM gas.
}
\label{fig03}
\end{figure}

Using the diffusive galactic model described in \S\ref{DiffusionGalacticModel}, we calculated the 
diffusive galactic spectrum from SNRs. From the proton and helium spectra and from the ISM gas distributions, 
we estimated the charged pion production. Using the calculated cross section for neutrino production, 
we calculated the flux of neutrinos at Earth coming from the galaxy bulge, for neutrino energy in the range $1~MeV-100~GeV$.
In Fig.\ref{fig02}, the total neutrino flux and the contribution of each kind of neutrino  are shown.  
In Fig.\ref{fig03}, the skymap of the total neutrino energy flux in Galactic latitude and longitude coordinates is shown.

\section{Acknowledgements}

CDD thanks ICN-UNAM for hosting a long stay and Universit\`a degli Studi di Milano for a PhD grant. GMT and JCD thanks PAPIIT/CIC-UNAM for support.

\bibliography{icrc1252}
\bibliographystyle{elsart-num}

\end{document}